\documentclass[fleqn,12pt,twoside]{article}
\usepackage{espcrc1}

\usepackage{graphicx}
\usepackage[figuresright]{rotating}

\newcommand{\AmS}{{\protect\the\textfont2
  A\kern-.1667em\lower.5ex\hbox{M}\kern-.125emS}}

\def\ltsim{\raise0.3ex\hbox{$<$\kern-0.75em\raise-1.1ex\hbox{$\sim$}}}
\def\gtsim{\raise0.3ex\hbox{$>$\kern-0.75em\raise-1.1ex\hbox{$\sim$}}}
\def\Tini{\langle T \rangle_{\rm ini}}
\def\Tmax{T_{\rm max}}

\title{On Hydrodynamical Description of Thermal Photons}

\author{S.~S.~R\"as\"anen \address[JYFL]{
                Department of Physics, PB 35 (YFL), 
                FIN-40014 University of Jyv\"askyl\"a, 
                Finland}\thanks{E-mail: sami.rasanen@phys.jyu.fi}}
\begin{document}

\maketitle

\begin{abstract}

The WA98 collaboration in the CERN SPS has reported an excess of photons 
over those originating from the decays of final hadrons in the lead-lead 
collisions. These photons can originate either from primary interactions 
of partons from colliding nuclei or from secondary interactions among 
produced particles. Photons produced in the secondary interactions, 
often called thermal photons, can be calculated by using thermal production 
rates and equilibrium hydrodynamics for the evolution of the expanding matter. 
I will review the main features of hydrodynamical studies for the WA98 data. 
The data can be reproduced both with or without a phase transition to the 
QGP, but high initial temperature, over the values predicted for the phase 
transition temperature, is required by the data. I will also show a prediction 
for the photon excess for central gold-gold collisions at the 
Brookhaven RHIC collider. In this prediction, the initial state for the 
hydrodynamical expansion is obtained from a perturbative QCD calculation.

\end{abstract}

\section{Introduction and Theoretical Framework}

Photons are suggested to be an ideal probe for the hot initial state 
of the relativistic heavy ion collisions, since, due to the large mean free 
path in a hadronic matter, they escape from the collision zone right after 
being emitted without any further interactions. Unfortunately, photons emitted 
at early stages of the collision are difficult to obtain experimentally, 
since the decays of the final state hadrons, especially neutral pions and 
etas, dominate the production of photons in the collision. So far only 
the WA98 collaboration at the CERN SPS has reported an excess of photons 
over those originating from decays of the final state hadrons 
\cite{WA98a,WA98b}. Photons are measured also by the PHENIX collaboration 
at Brookhaven RHIC collider, but more statistics is needed to reach 
conclusions \cite{Klaus}.

In a hydrodynamical description of relativistic heavy ion collision the 
secondary interactions are assumed to lead to thermalization of the fireball. 
With this assumption, the evolution of the system can be described by 
equilibrium hydrodynamics and transverse momentum spectrum of the 
{\em thermal photons} emitted during the expansion can be calculated if 
thermal emission rate in the matter is known. Experimentally measured 
{\em direct} photon spectrum has a contribution from primary interactions 
as well \cite{Dumitru0}, referred as pQCD (or prompt) photons in following.

Several groups \cite{Oma,SS,PP,Ch,Alam} have performed hydrodynamical studies 
to explain WA98 data. The authors of \cite{Oma} use a (2+1)-dimensional 
hydrodynamical code that takes into account also the finite longitudinal size 
of the initial collision zone, whereas the studies in \cite{SS,PP,Ch,Alam} are 
based on the boost-invariant approximation. In the boost-invariant scenario 
very hot initial state \cite{SS}, considerable initial radial 
velocity \cite{PP,Ch,Alam} or modification of hadron masses \cite{Alam} 
is needed to explain the WA98 data. I have made a choice to omit 
discussion on modifications of hadron masses \cite{Alam} and also chosen 
references \cite{SS,PP} to represent boost-invariant studies in this note.

Important improvements have been obtained in a perturbative calculation of 
the thermal emission rate of photons from the thermalized QGP 
\cite{Zaraket,Arnold2,Arnold}. In hadron gas mesonic processes 
$\pi\pi\rightarrow\rho\gamma$ and $\pi\rho\rightarrow\rho\gamma$, described 
with a pseudo-vector Lagrangian, are included in the photon emission rate 
\cite{Kapusta}. The $\pi\rho$ scattering channel gets a large contribution 
from the interaction with the a$_1$ axial meson: 
$\pi\rho\rightarrow{\rm a}_1\rightarrow\rho\gamma$ \cite{Xiong}. 
Parametrizations of the emission rates are provided by the authors of 
\cite{Xiong,Nadeau}. The effects of baryons are not included in the rates.

\section{Details of Hydrodynamics and Results}

In section \ref{SPS} results of three different groups \cite{Oma,SS,PP} are 
compared to the WA98 data \cite{WA98a,WA98b}. Groups \cite{SS,PP} use 2-loop 
photon emission rates in QGP \cite{Aurenche,Steffen} while in \cite{Oma} 
resummed rate \cite{Arnold}, complete in order $\alpha_s$, is used. In 
works \cite{Oma,SS} contribution of the pQCD photons is taken from \cite{WW}. 
The authors of \cite{PP} present their own estimate for the pQCD photons.

In section \ref{RHIC} a prediction for direct photon spectrum in Au$+$Au 
collisions with $\sqrt{s}=200$ AGeV is given. Results for hadronic 
observables in this study, where the initial state is based on perturbative 
QCD $+$ saturation model, can be found in \cite{Oma2,Oma3}. For this energy, 
hard pQCD photons from the primary interactions are studied in \cite{Dumitru}.

The following discussion will concentrate on the role of the initial state in 
hydrodynamical models, since especially high transverse momentum $k_t$ 
part of the thermal photon spectrum is sensitive to the initial conditions. 
In particular, temperature $T$ in the initial state plays a big role due to
the Boltzmann suppression $e^{-k_t/T}$ in the emission rates.

\subsection{Comparison with the WA98 Data} \label{SPS}

In boost-invariant hydrodynamical studies one must give the initial 
energy density $\epsilon(r,\tau_0)$ and velocity $v(r,\tau_0)$ distributions 
in the transverse plane at fixed formation (thermalization) time $\tau_0$, 
that are the initial conditions for the hydrodynamical evolution. Let us start 
with a case $v(r,\tau_0)\equiv0$ for all $r$. For the shape of the initial 
energy density $\epsilon(r,\tau_0)$ various choices can be found in the 
literature. For example, it can be proportional to wounded nucleon 
distribution \cite{SS}, Woods-Saxon distribution \cite{PP} or 
nuclear overlap function \cite{Oma2}. Within these choices, the formation time 
$\tau_0$ remains as the main uncertainty in the boost-invariant picture.
At SPS energies, $\tau_0$ is expected to be $\sim1$ fm/c based on the 
geometrical argument, that it takes a time $\sim 2R_A/\gamma$, where 
$R_A$ is the nuclear radius and $\gamma$ the Lorentz gamma-factor, for the 
colliding nuclei to pass through each other. Fixing $\tau_0=1$ fm/c leaves 
only the normalization of the initial energy density as a free parameter, that 
can be fixed from experimentally measured hadron spectra.

\begin{figure}[t]
\begin{minipage}[t]{75mm}
  \begin{center}
    \includegraphics[height=5.5cm]{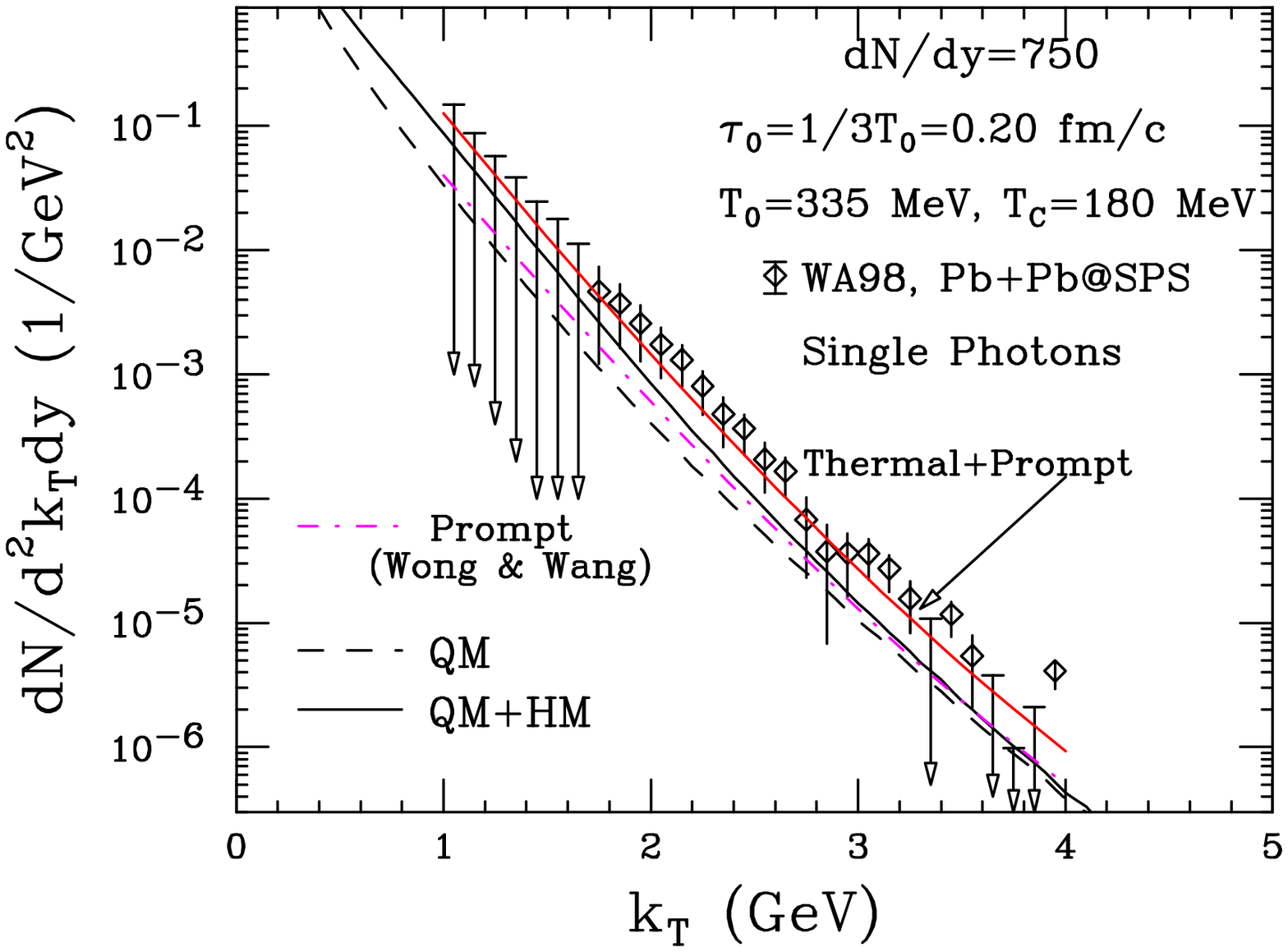}
  \end{center}
 \vspace{-1.3cm}
 \caption{Direct photon production in ref. \cite{SS} with the WA98 data. 
          QM refers to emission in QGP and QGP part of mixed 
          phase, HM likewise for hadronic matter. $T_c$ is a critical 
          temperature in EoS, $\tau_0$ thermalization time and $T_0$ 
          average initial temperature.}
 \label{fig:calcutta}
\end{minipage}
\hspace{\fill}
\begin{minipage}[t]{75mm}
  \begin{center}
    \includegraphics[height=5.5cm]{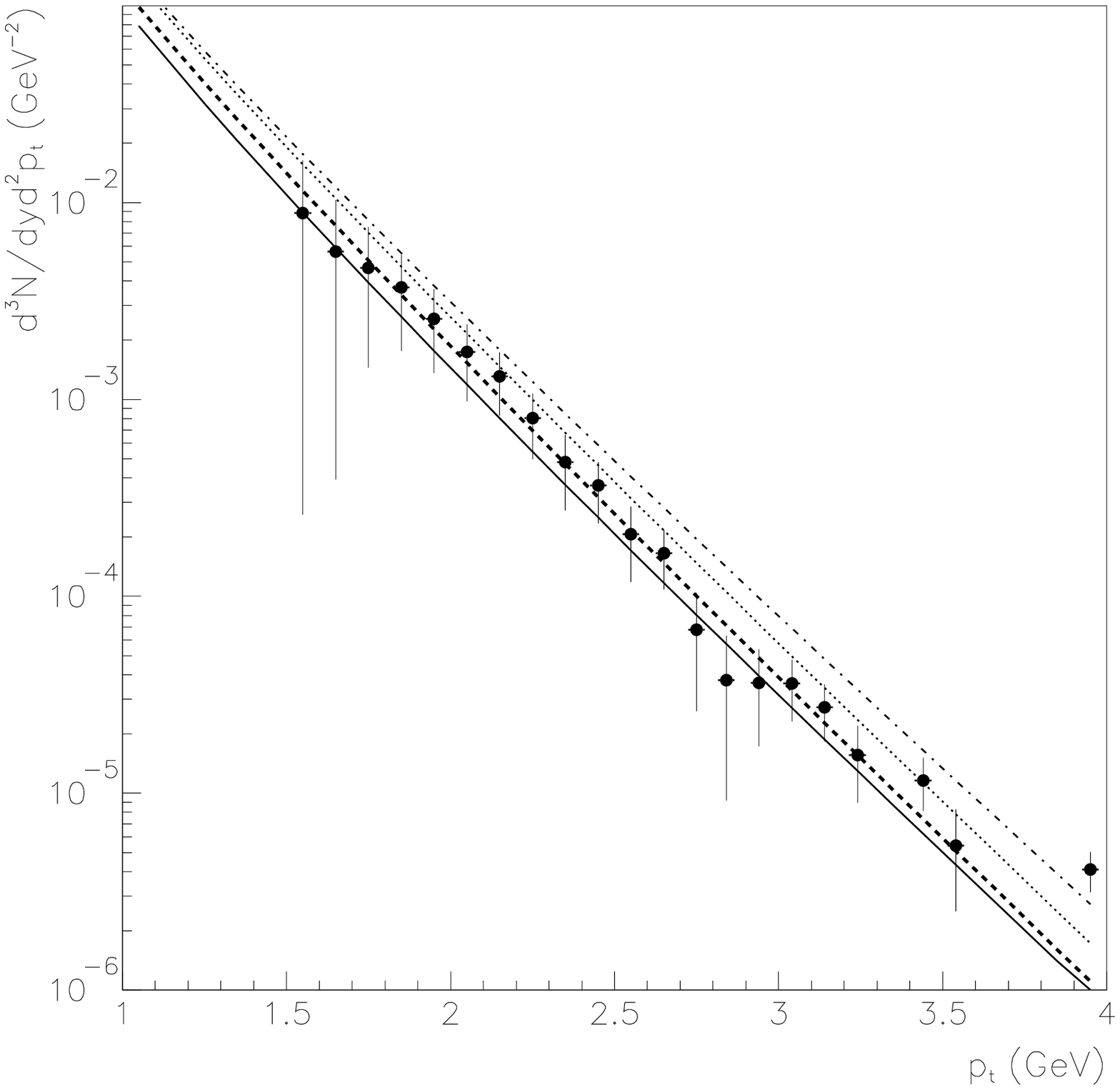}
  \end{center}
 \vspace{-1.3cm}
 \caption{Direct photon yield calculated for EoS with phase transition 
          and linear initial velocity distribution in ref. \cite{PP} with 
          the WA98 data. Initial parameters for different curves in the 
          figure are presented in the table \ref{tab:moscow}. Estimate 
          for pQCD (prompt) photons is included to the curves.}
 \label{fig:moscow}
\end{minipage}
\end{figure}

Studies by the authors of \cite{SS} reveal a problem in this simple approach. 
If initial time is chosen to be of the order 1 fm/c, the initial 
state is way too cool to reproduce high transverse momentum, $k_t$, part of 
the spectrum as measured by the WA98 collaboration. Since decreasing of 
initial time considerably raises the temperature in the initial 
state\footnote{If radial expansion is omitted, 
$T(r,\tau)=T(r,\tau_0)\times(\tau/\tau_0)^{-1/3}$ for massless ideal gas.} 
and final state hadron spectra are fairly insensitive 
to small changes of $\tau_0$, the data seems to suggest very rapid 
thermalization and hot initial state within the boost-invariant scenario. 

Figure \ref{fig:calcutta} shows the main results in \cite{SS}. The data is 
well reproduced with initial time $\tau_0=0.2$ fm/c which corresponds to an 
average initial temperature $\Tini=335$ MeV in their choice of the initial 
state. This temperature is considerably higher than the expected critical 
temperature $T_c=150-180$ MeV from recent lattice calculations \cite{Karsch} 
and hence the authors of \cite{SS} conclude that this confirms the formation 
of the QGP. However, in my opinion, one should consider the initial time 
$\tau_0=0.2$ fm/c carefully, although the WA98 data seem to favor it. The 
geometrical argument above, on the transit time of the nuclei, suggests that 
at such a short initial time primary collisions are still taking place, and 
only part of the matter has been produced \cite{KRR}.

Let us turn to discuss effects from non-zero initial radial velocity. The 
authors of \cite{PP} have studied the effects of radially increasing initial 
velocity $v(r,\tau_0)=(r/R_A)\Theta(R-r)V^{\rm max}$, where $V^{\rm max}$ is 
treated as a free parameter. They have performed a simultaneous $\chi^2$ 
analysis of $\pi^0$ and $\gamma$ spectra measured by the WA98 collaboration 
\cite{WA98a} to find the best combination of the thermalization time $\tau_0$ 
and the velocity parameter $V^{\rm max}$. Their results \cite{PP} with a 
phase transition to QGP are presented in figure \ref{fig:moscow}. Curves in 
the figure \ref{fig:moscow} correspond to the choices of the initial 
parameters presented in the table \ref{tab:moscow}.
\begin{table}[htb]
\caption{Choice of the initial parameters in \cite{PP}.}
\label{tab:moscow}
\begin{center}
\begin{tabular}{|c|c|c|c|}
\hline
$\Tini$ [MeV] & $\tau_0$ [fm/c] & $V^{\rm max}$ [c] 
                                 & Curve in the figure \ref{fig:moscow} \\
\hline
180 & 2.0 & 0.40 & solid \\
200 & 1.5 & 0.38 & dashed \\
230 & 1.0 & 0.29 & dotted \\
250 & 0.7 & 0.26 & dash-dotted \\
\hline
\end{tabular}
\end{center}
\end{table}
The results show clear interplay between the initial temperature and the 
strength of the initial radial velocity: thermal photons with large $k_t$ are 
emitted at first moments of the evolution when the temperature is high. 
Lowering of initial temperature, or equivalently increasing the thermalization 
time $\tau_0$, requires an introduction of initial radial velocity, which 
boosts the emitted photons to larger $k_t$. With this compensation, initial 
temperature can be lowered below 200 MeV and the thermalization time 
rises to values $\tau_0\gtsim1$ fm/c. These values of the temperature are 
quite close to the critical temperature and hence the authors of \cite{PP} 
conclude, that, considering uncertainties of the model, formation of QGP or 
its absence cannot be distinguished.

The origin and the magnitude of the initial radial velocity is a somewhat 
uncertain. The authors of \cite{PP} explain its origin ``as a result of 
significant radial energy gradient coming from the shapes of the colliding 
nuclei''. They do not present a (quantitative) model that could explain how 
this leads to a collective motion into the radial direction at initial time 
$\tau_0$. In this sense, I would consider $V^{\rm max}$ to be an extra fitting 
parameter in the model. In any case, initial radial velocity can not be large. 
The authors of \cite{Oma} have made a boost-invariant calculation with initial 
radial velocity profile taken from \cite{PP}. They found a good agreement with 
the photon data by choosing initial parameters $\tau_0=1.0$ fm/c and 
$V^{\max}=0.30$, but the resulting $\pi^0$ spectrum clearly overshoots the 
experimental one. It should be noted that the initial energy density profile 
--- normalization is fixed from multiplicity in both cases --- and the 
equation of state are different in these two studies, and in \cite{Oma} 
chemical and kinetic freeze-out temperatures are the same. These differences 
may explain the discrepancy on the $\pi^0$ spectrum.

The authors of \cite{Oma} used a (2+1)-dimensional hydrodynamical code, where 
also the longitudinal extend of the fireball is finite and the longitudinal 
velocity evolves dynamically from a given initial state. Details of this model 
can be found in references \cite{Pasi1,Pasi2,Pasi3}. My first remark is, that 
there is no uniquely defined initial time, when the assumption of 
boost-invariant flow is relaxed. In the boost-invariant case the source is 
pointlike in the $zt$-plane and one can trace all the particles back to the 
collision point. This cannot be done when the longitudinal flow profile of the 
system is not that of the scaling flow, $v_z = z/t$. To estimate the 
thermalization time one can relate a timescale to the longitudinal extent by 
considering the collision geometry. In the model \cite{Pasi2} the initial 
length of the system was chosen to be 3.4 fm. Since $\gamma\sim10$ at SPS 
energies, the system is roughly 1.3 fm thick when the colliding nuclei 
overlap and it would take $\sim1$ fm/c for the system to reach longitudinal 
extent of 3.4 fm \cite{PH}. The rapidity distribution of initial 
energy density is assumed to be Gaussian. The energy per unit transverse area, 
$e(r)=\int dz T^{00}(r,z)$, is assumed to equal the energy per unit transverse 
area of the incoming nucleons. To convert this distribution to spatial energy 
density, one must define initial velocity profiles. Initial conditions, 
chosen to reproduce the observed hadron spectra, depend also on the equation 
of state (EoS). In \cite{Oma} EoS~H describes a hadron resonance gas without 
phase transition and EoS~A has a first order phase transition to QGP. Initial 
radial velocity is set to zero. For the longitudinal velocity profile two 
different choices were studied \cite{Pasi3}; in the case called IS~1 the 
longitudinal flow rapidity is assumed to increase linearly with the distance 
$z$. This gives a large peak, by a factor of two over the Bjorken estimate, to 
the initial energy density. In the other choice, IS~2, the $z$ dependence is 
nonlinear and the resulting energy density profile is almost flat. 

\begin{figure}[t]
\begin{minipage}[t]{75mm}
  \begin{center}
    \includegraphics[height=6.3cm]{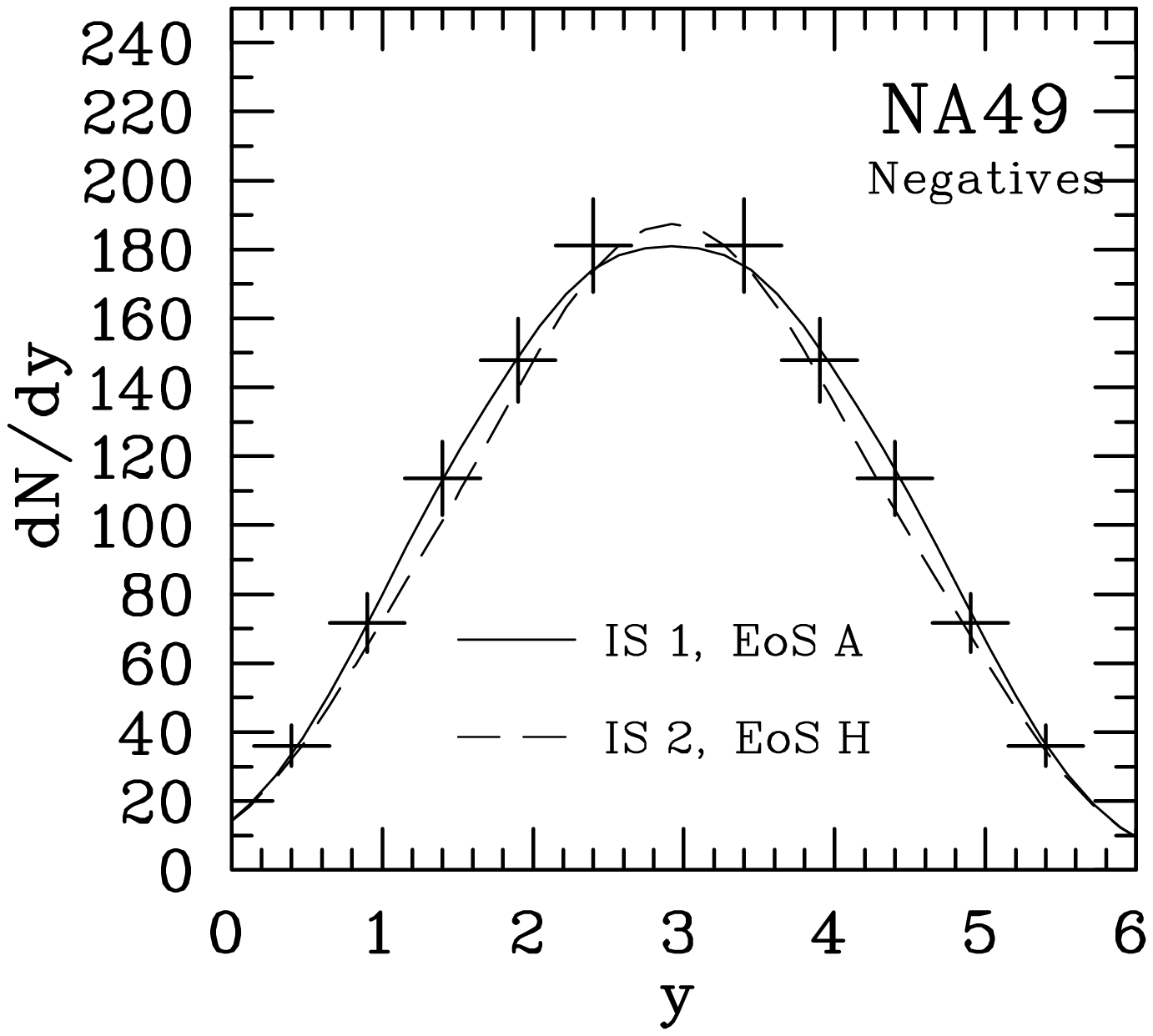}
  \end{center}
 \vspace{-1.3cm}
 \caption{Number density $dN/dy$ of negatively charged hadrons as a 
          function of rapidity $y$ with the NA49 data \cite{Oma,Pasi3}.}
 \label{fig:rapid}
\end{minipage}
\hspace{\fill}
\begin{minipage}[t]{75mm}
  \begin{center}
    \includegraphics[height=6.3cm]{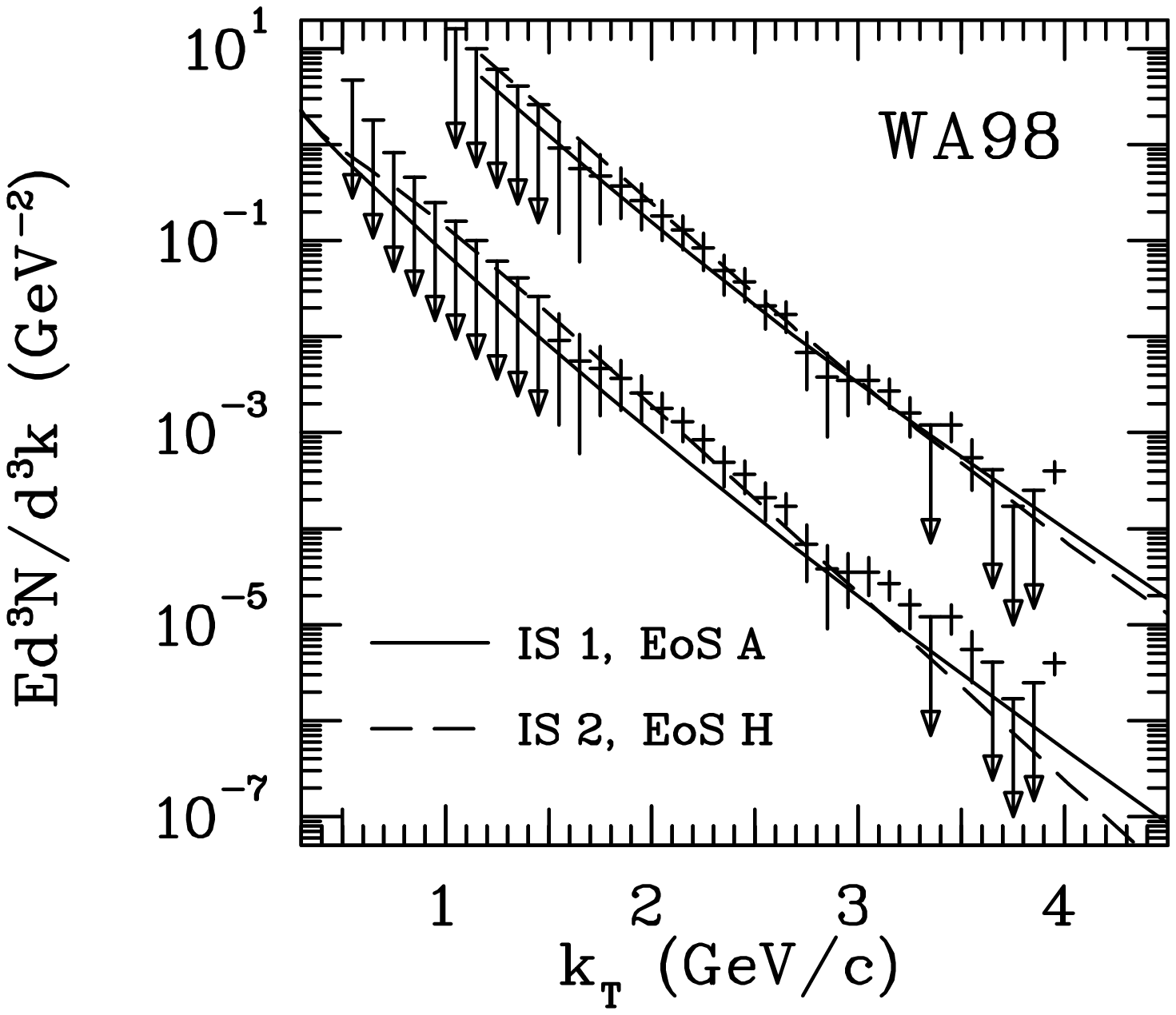}
  \end{center}
 \vspace{-1.3cm}
 \caption{Thermal photon emission with the contribution of pQCD photons 
(upper set, scaled by factor of 100) and without (lower set) in 
ref. \cite{Oma}. Solid line show the results with QGP formation and dashed 
line in purely hadronic scenario.}
 \label{fig:phot}
\end{minipage}
\end{figure}

Combinations IS~1 + EoS~A and IS~2 + EoS~H reproduce {\em simultaneously} the 
NA49 data for the rapidity distributions of hadrons \cite{NA49}, and the WA98 
data for neutral pions and direct photons \cite{WA98a}. Figure \ref{fig:rapid} 
shows the results for measured rapidity density of negative charged hadrons, 
comparison to $\pi^0$ spectrum can be found in \cite{Oma} and figure 
\ref{fig:phot} shows the results for direct photons. Maximum and average 
temperatures in $z=0$ in the initial state are given in the table 
\ref{tab:jkl}. 

\begin{table}[htb]
\caption{Maximum and average initial temperature in \cite{Oma}.}
\label{tab:jkl}
\begin{center}
\begin{tabular}{|c|c|c|}
\hline
  & IS~1 + EoS~A & IS~2 + EoS~H \\
\hline
$\Tmax$ [MeV] &  325   & 245 \\
$\Tini$ in $z=0$ [MeV] &  255   & 213 \\
\hline
\end{tabular}
\end{center}
\end{table}

At the given temperature, the thermal emission rate of photons is larger in 
hadronic matter, which makes IS~2 with lower initial temperature more suitable 
for purely hadronic scenario. Choosing IS~1 $+$ EoS~A leads to a high initial 
temperature, but gives a good agreement also with the high $k_t$ part of the 
measured photon spectrum, as shown in the figure \ref{fig:phot}. Despite the 
$\sim50$ MeV difference in the initial temperature, later evolution 
is so similar in these two choices for the initial state, that hadronic 
spectra are equally well reproduced. These observations reflect the ambiguity 
in the choice of the initial state.

After pQCD photon spectrum from the ref. \cite{WW} is added to the thermal one 
(upper set of curves, scaled by a factor of 100, in the figure 
\ref{fig:phot}), the authors of \cite{Oma} find that the WA98 data can 
be reproduced both with or without a phase transition to QGP. However, the 
maximum (average) temperature 245 (213) MeV in the initial state with hadron 
gas is fairly high, and clearly over the expected phase transition temperature 
150$-$180 MeV \cite{Karsch}. Also the shape of the solid curve in the figure 
\ref{fig:phot}, which corresponds to the results from IS~1 $+$ EoS~A with the 
phase transition, seems to follow the data better.

\subsection{Thermal Photons at RHIC Collider} \label{RHIC}

References \cite{Oma2,Oma3} show results for hadronic observables from 
boost-invariant hydrodynamics, when the initial state is obtained from 
a pQCD + saturation model. With this model for the initial state, only free 
parameter in the hydrodynamical description is the decoupling temperature. 
The initial time and energy density, including normalization, can be 
calculated from the model when the center of mass energy of the collision, 
$\sqrt{s}$, mass number of the colliding nuclei, $A$, and the experimental 
centrality cut are given. Construction of the initial state is explained in 
\cite{Oma2}, where also results for global observables (e.g. multiplicity) 
are shown. In \cite{Oma3} the effect of the decoupling temperature is 
discussed and results are compared to hadron spectra measured by the PHENIX 
collaboration \cite{phe1,phe2}. Following results for thermal photons are 
calculated with initial parameters $\sqrt{s}=200$ AGeV and $A=197$, which 
lead to $\tau_0=0.17$ fm/c, $\sigma\langle E_T\rangle=86.33$ mbGeV and 
$A_{\rm eff}=178$ for a 6 \% most central collisions. For details see 
\cite{Oma2}.

The initial time is determined from the saturation momentum scale, 
$p_{\rm sat}$, of the perturbative minijet calculation \cite{EKRT}. This 
leads to a very short initial time $\tau_0=1/p_{\rm sat}=0.17$ fm/c compared 
to values $\gtsim0.6$ fm/c, that are used in many recent hydrodynamical 
studies at RHIC energies, for a review see e.g. \cite{Pasi4}. The short 
initial time $\tau_0=0.17$ fm/c is consistent with the geometrical argument 
presented in section \ref{SPS}. The produced minijet system also looks thermal 
from the point of view of the number of the partons and the energy per 
particle at the initial time \cite{EKRT}. Starting hydrodynamics at 
$\tau_0=0.17$ fm/c can be considered to give an upper limit for the initial 
temperature and the thermal photon production.

\begin{figure}[t]
\begin{minipage}[t]{75mm}
  \begin{center}
    \includegraphics[width=8.0cm]{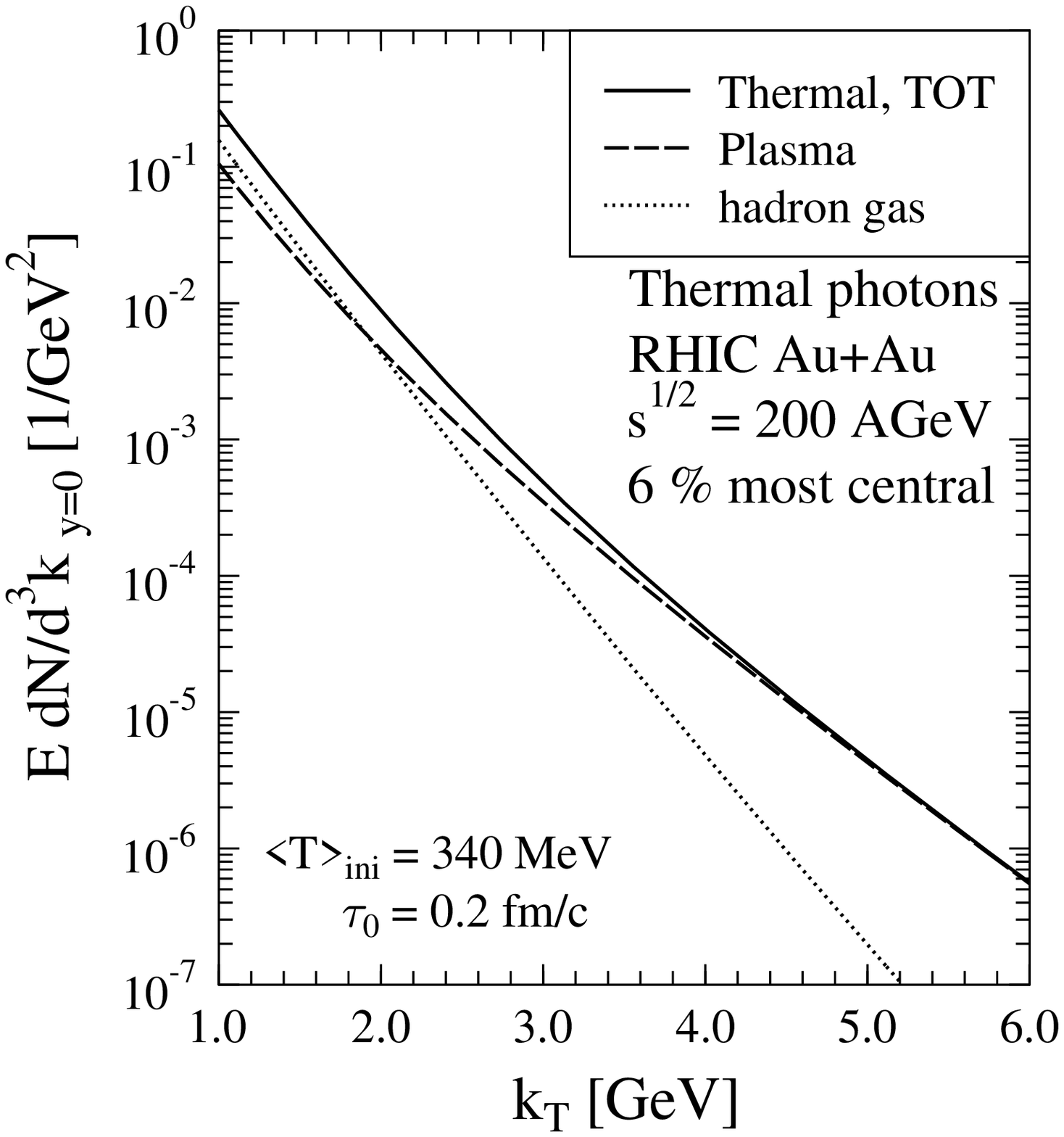}
  \end{center}
 \vspace{-1.3cm}
 \caption{Thermal photon spectrum in Au+Au collision with $\sqrt{s}=200$ AGeV. 
Plasma refers to the contribution from QGP and QGP part of the mixed phase.
Also hadron gas includes emission from hadronic part of the mixed phase.}
 \label{fig:coctail}
\end{minipage}
\hspace{\fill}
\begin{minipage}[t]{75mm}
  \begin{center}
    \includegraphics[width=8.0cm]{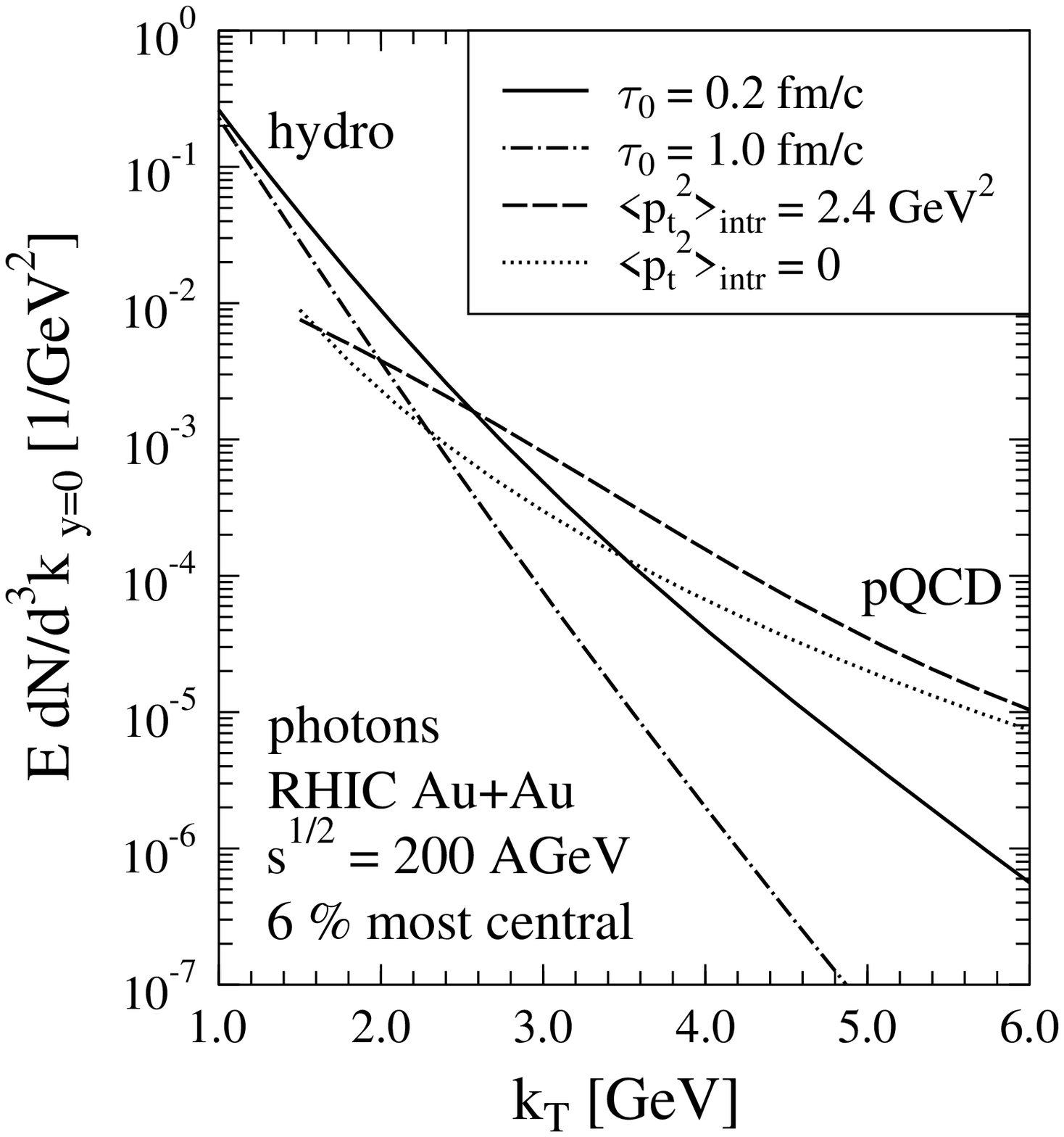}
  \end{center}
 \vspace{-1.3cm}
 \caption{Effect of initial time $\tau_0$ to the thermal photon emission. 
  Spectra of pQCD photons, with two different values for intrinsic 
  transverse momentum $\langle p_t^2\rangle$, are from work by A.~Dumitru 
  {\em et al.} \cite{Dumitru0,Dumitru}. Initial temperatures are in the table 
  \ref{tab:rhic}.}
 \label{fig:rhic}
\end{minipage}
\end{figure}

The solid line in the figure \ref{fig:coctail} shows the transverse momentum 
spectrum of thermal photons. The dashed line shows contribution from 
quark-gluon plasma and QGP part of the mixed phase, and the dotted line 
likewise from hadron gas. The plasma contribution dominates the thermal 
emission for $k_t>3$ GeV. Comparing results from SPS and RHIC energies 
one can see roughly a factor of 100 increase in thermal emission at $k_t=4$ 
GeV, but the difference is less than order of magnitude at $k_t=1$ GeV, 
which reflects a clear increase of the maximum temperature.

To analyze the effect from initial temperature (thermalization time) I 
have scaled the initial state from $\tau_0=0.17$ to 1.0 fm/c with 
longitudinal Bjorken expansion so that entropy in unit rapidity does not 
change. Neglecting transverse  expansion before $\tau=1.0$ fm/c is not 
physical, but in this way one can define a cooler initial state in such 
a way that the results for hadronic observables do not change much. Table 
\ref{tab:rhic} shows the changes in the maximum and average values of the 
initial temperature and energy density. Effect on the thermal photon 
spectrum is 
\begin{table}[htb]
\caption{Maximum and average initial temperatures and energy densities 
         in Au+Au collisions with $\sqrt{s}=200$ AGeV.}
\label{tab:rhic}
\begin{center}
\begin{tabular}{|l||c|c|c|c|}
\hline
       & $\Tmax$ [MeV] & $\Tini$ [MeV] & $\epsilon_{\rm max}$ [GeV/fm$^3$] & 
                     $\langle\epsilon\rangle_{\rm ini}$ [GeV/fm$^3$] \\
\hline
$\tau_0=0.17$ fm/c & 580 & 340 & 208 & 60  \\
$\tau_0=1.0$  fm/c & 320 & 220 &  20 & 6.8 \\
\hline
\end{tabular}
\end{center}
\end{table}
presented in the figure \ref{fig:rhic}.

Increasing initial time from $\tau_0=0.17$ to 1.0 fm/c reduces the thermal 
emission by a factor of 20 at $k_t=4$ GeV. Unfortunately, the amount of 
thermal photons in this region may be hard to resolve experimentally,
because pQCD photons from primary interactions may dominate the direct 
photon spectrum there. The analysis of the pQCD photons can be found in 
\cite{Dumitru}. In \cite{Dumitru} the centrality cut is 10 \% instead of 6, 
so the pQCD results presented in the figure \ref{fig:rhic} are multiplied with 
a factor of 1.1 \cite{Dumitru2}. Results for two different values of 
the intrinsic transverse momentum, $\langle p_t^2\rangle=0$ and 2.4 GeV$^2$, 
are given. 

These theoretical results on excess photons at RHIC can be summarized as an 
upper limit given by the sum of the solid and dashed lines and a lower limit 
given by the sum of the dotted and dash-dotted lines. Uncertainties 
are fairly large, but in any case one should see a change in the slope in the 
region $k_t\sim2.0-3.5$ GeV, above which the pQCD photons start to dominate 
the spectrum. In principle a very clean data around $k_t\sim2$ GeV could fix 
the value of the initial time, but one should keep in mind the possible 
effects from initial transverse velocity and longitudinal geometry discussed 
in the section \ref{SPS}. These effects may be smaller at RHIC energies, 
because rapid thermalization means shorter time interval to build up the 
initial transverse velocity. Also the rapidity range is broader at RHIC, and 
hence scaling hydrodynamics should work better.

\section{Conclusions}

I have reviewed hydrodynamical studies to explain the WA98 excess photon 
spectrum. The role of the initial state and the uncertainty in the values of 
initial temperature were emphasized in the discussion. The data can be 
explained with or without QGP phase transition, but high initial 
temperatures, well above the predicted critical temperature, are 
favored. In the case without boost-invariance the shape of the spectrum is 
closer to the data, when a phase transition is assumed. Still, both 
theoretical and experimental uncertainties are somewhat too large to rule 
out either alternative.

Using boost-invariant hydrodynamics with an initial state calculated from a 
pQCD + saturation model, theoretical estimates of the upper and lower 
limits for the thermal photons were given and compared with the results 
for the pQCD photons \cite{Dumitru0,Dumitru}.
\newline\newline
{\bf Acknowledgments:}
\newline\newline
The author is grateful to P.~V.~Ruuskanen for discussions, P.~Huovinen for 
comments and making the figures 3 and 4, and A.~Dumitru and L.~Gerland for 
sending their results for pQCD photons. The author would like to thank the 
GRASPANP graduate school, funded by the Ministry of Education and the 
Academy of Finland, for financial support.

\end{document}